\documentclass[11pt]{article}
\usepackage{moriondav,epsfig}
\usepackage{wrapfig}

\def\Journal#1#2#3#4{{#1} {\bf #2}, #3 (#4)}

\def\PRL{\em Phys. Rev. Lett.}

\newcommand{\met}     {\mbox{$\not\!\!{E_T}$}}
\def\gsim{\mathrel{\rlap{\raise.4ex\hbox{$>$}} {\lower.6ex\hbox{$\sim$}}}}
\def\lsim{\mathrel{\rlap{\raise.4ex\hbox{$<$}} {\lower.6ex\hbox{$\sim$}}}}

\begin{document}
\vspace*{4cm}
\title{TOP QUARK PRODUCTION CROSS SECTION AT THE TEVATRON}

\author{ IA IASHVILI \\
(For the CDF and D\O\ Collaborations) }

\address{Department of Physics, University of California,
Riverside, CA 92521, USA}

\maketitle\abstracts{
Preliminary results obtained by the CDF and D\O\ Collaborations on the
top quark pair production cross-section in $p\bar{p}$ collisions at
a center-of-mass energy of $\sqrt{s}$=1.96~TeV are presented.
The measurements are obtained using various final states from
top quark pair production and decays, and are based on data collected 
during years 2002-2004 of the Tevatron Run~II.}

\section{Introduction}

Since the discovery of the top quark 
by the CDF and D\O\ Collaborations~\cite{discovery}, 
Tevatron remains the world's only experimental facility where it's properties 
can be investigated. 
In $p\bar{p}$ collisions 
top quarks are dominantly produced in pairs via strong interaction. 
Depending on the decay modes of the $W$-bosons, 
one expects three distinctive event topologies from the 
$t\bar{t}\rightarrow W^+bW^-\bar{b}$ signal:
a di-lepton final state with 2 high-$p_T$ isolated leptons, 2 jets,
and missing $E_T$ (\met) from escaping neutrinos; a lepton+jets final state
with  a high-$p_T$ lepton, 4 jets and \met; and an all-jets final
state consisting of 6 jets. 
While the di-lepton channel gives the cleanest signal, it amounts to only 5\% of the
$t\bar{t}$ sample (considering only e and $\mu$), whereas the 
all-jets state constitutes 44\% of the total fraction, but suffers 
from a huge QCD multijet background. The lepton+jets final state, 
forming 30\% of $t\bar{t}$ events, is the 
best compromise between purity and the statistics.

At the Tevatron the top quark can also be produced singly, via 
electroweak interaction
This mechanism has not been observed yet, and remains
subject of active searches at the Tevatron~\cite{single_top}.

\begin{figure}[t]
\vskip -2mm
\centerline{
\includegraphics[bb=10 22 570 485,
width=72mm,height=65mm]{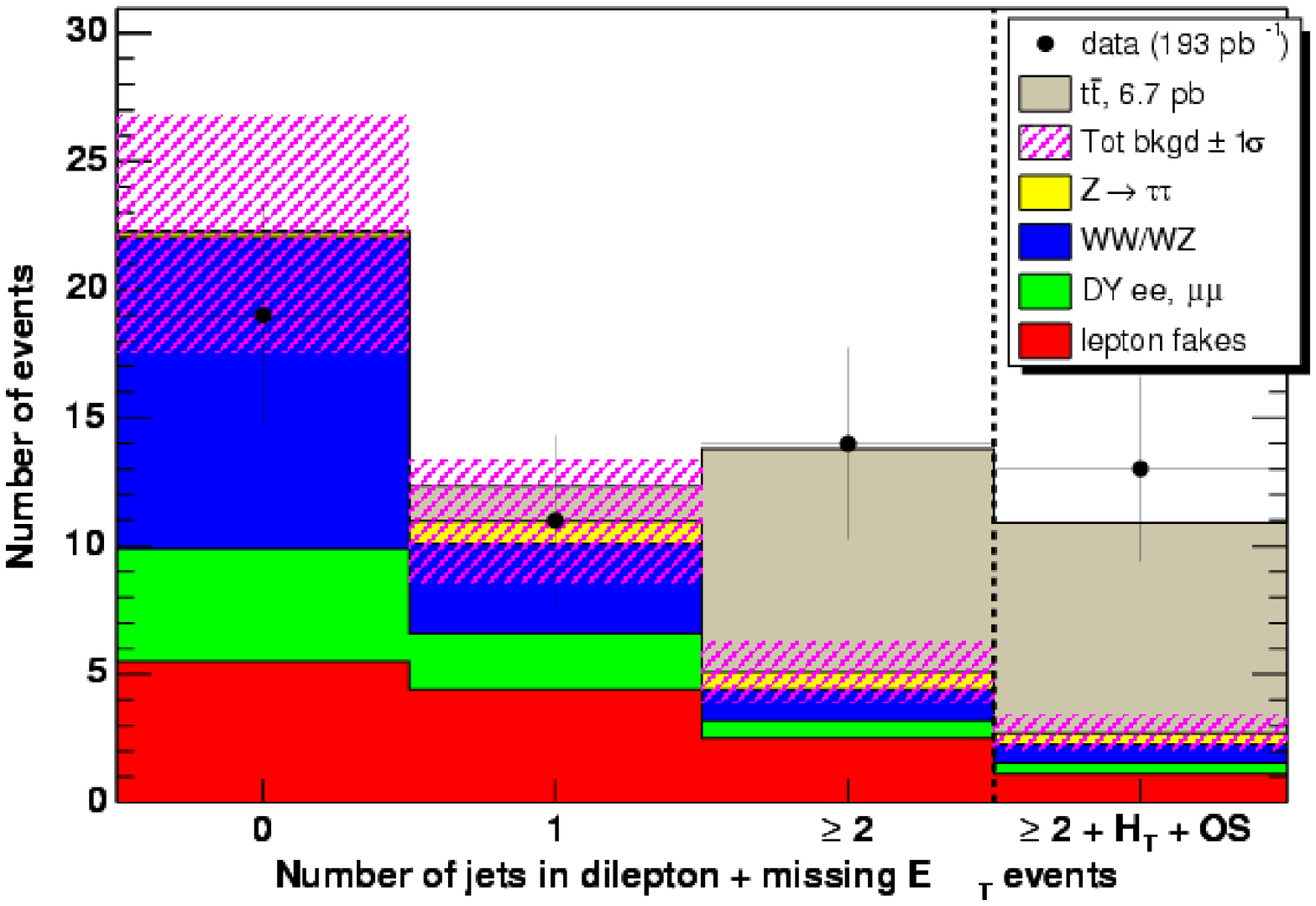}
\hskip 10mm
\includegraphics[bb=10 10 570 650,
width=72mm,height=65mm]{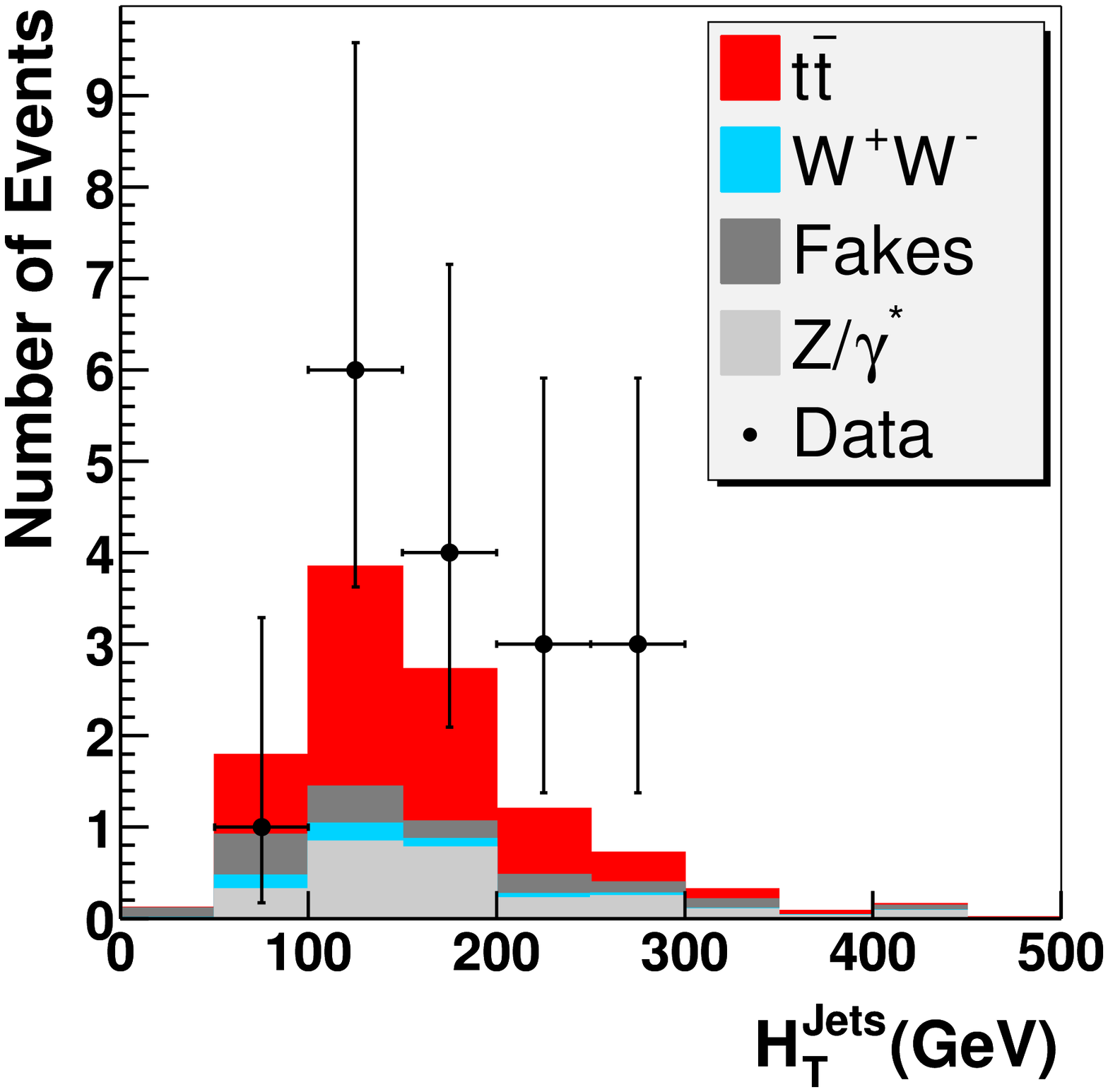}
}
\vskip -55mm
{\hskip 8mm} a) {\hskip 83mm} b)
\vskip 48mm
\caption{a)
Jet multiplicity distribution obtained by CDF in di-lepton events. 
The last bin corresponds to events with $\ge$2 jets, $
H_T>200$~GeV and opposite sign leptons;
b) Distribution of the sum of jet transverse momenta in di-lepton
events obtained by D\O.
\label{fig:dileptons}}
\vskip -3mm
\end{figure}

\section{Di-lepton channel}

An unique signature of $t\bar{t}$ arises when both W's decay leptonically
producing a pair of high-$p_T$ leptons, large $\met$ and two high-$p_T$
jets. The potential physics backgrounds to such a signature
come from $W^+W^-\rightarrow \ell^+\ell^-$ and 
$Z/\gamma^*\rightarrow \tau^+\tau^-$ events.
Each of these can be produced with $\ge$2 associated jets
and constitute significant background. 
Instrumental backgrounds arise from mismeasured $\met$ 
in 
$Z/\gamma^*\rightarrow \ell^+\ell^-$ production and from 
fake lepton in $W+jets$ events.


\begin{figure}[b]
\vskip -19mm
\centerline{
\includegraphics[bb=10 -10 590 600,width=55mm,height=54mm]
{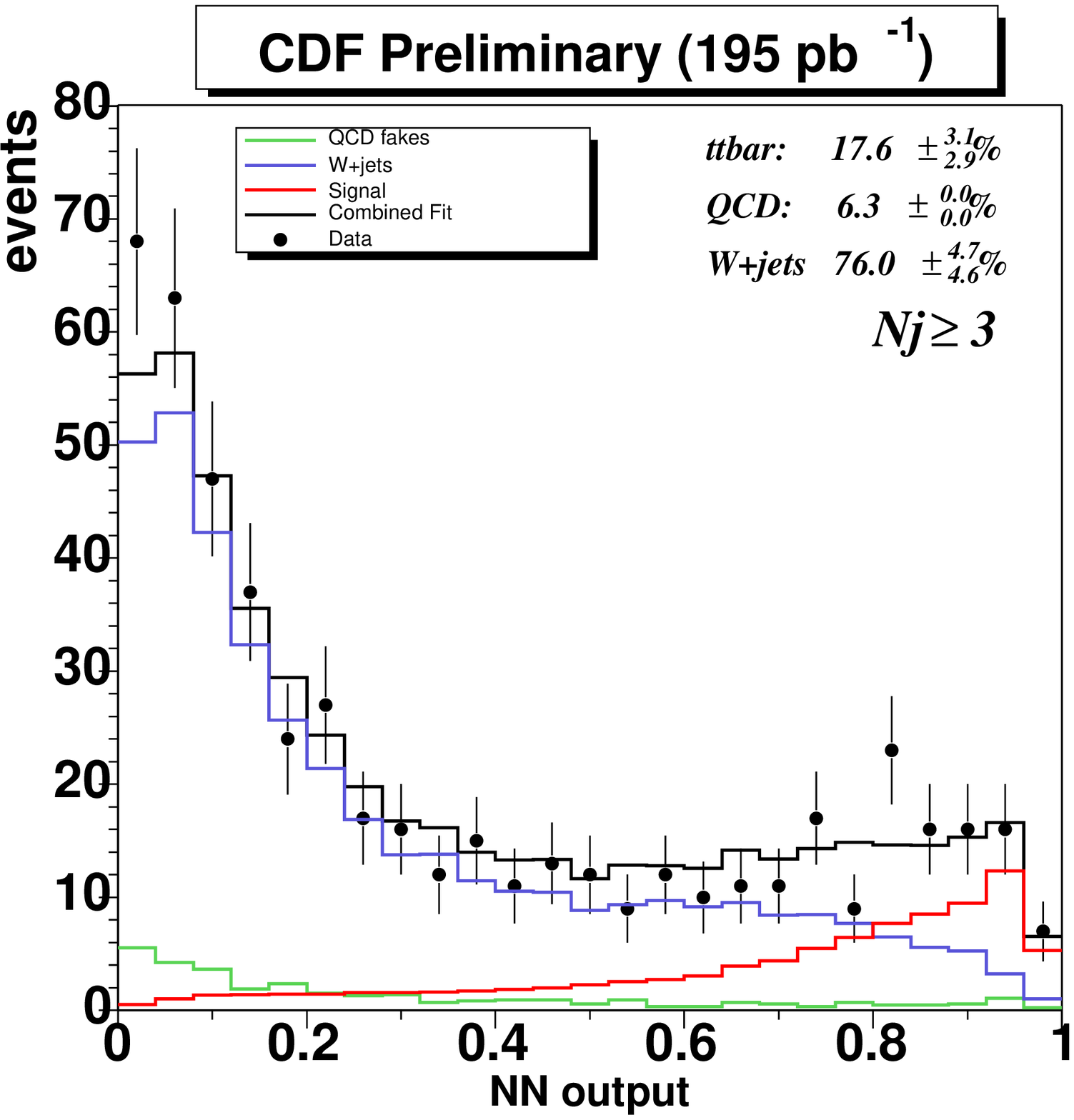}
\hskip -7mm
\includegraphics[bb=10 70 550 615,width=65mm,height=68mm]
{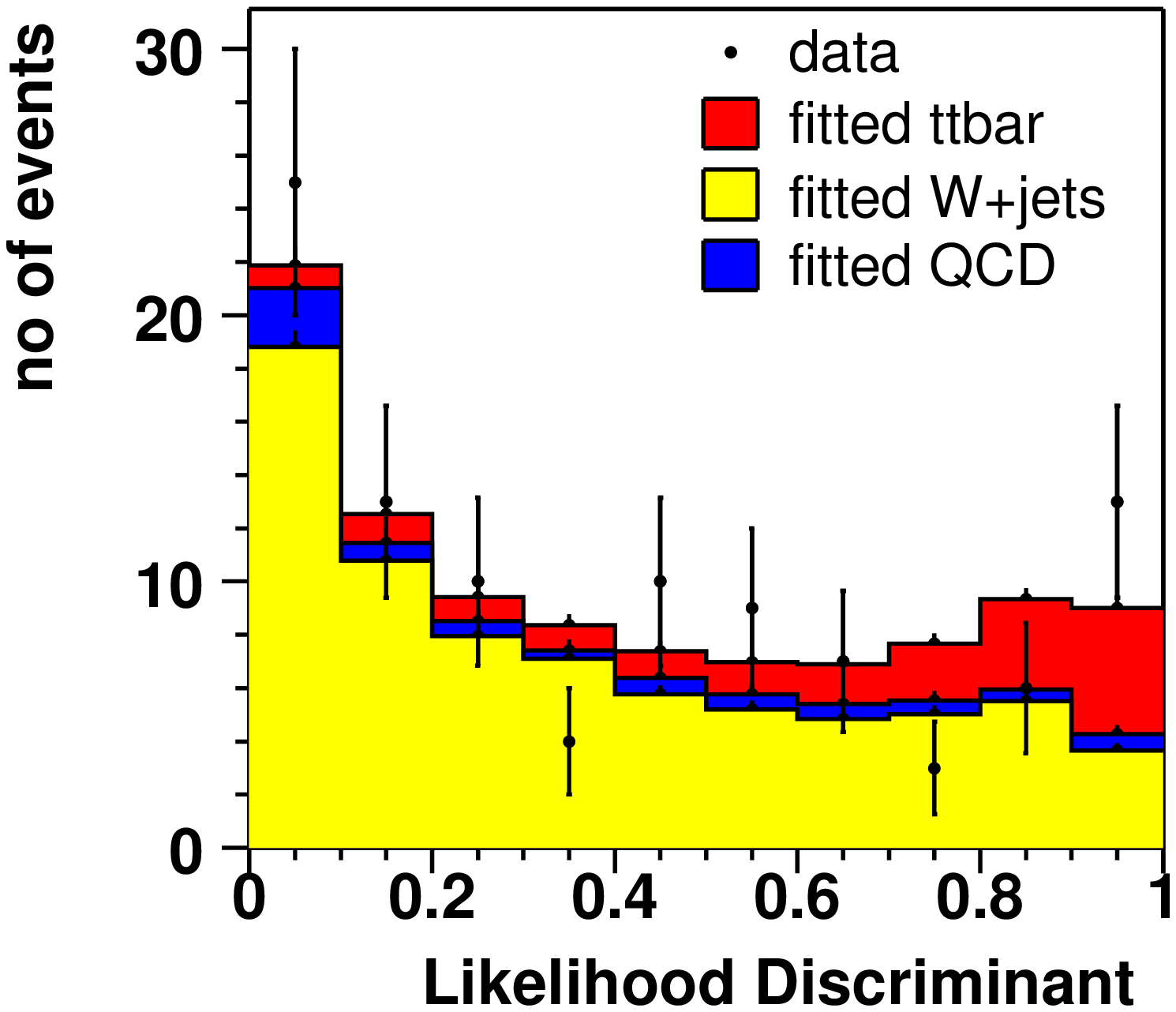}
\hskip -8mm
\includegraphics[bb=10 70 570 650,width=58mm,height=59mm]
{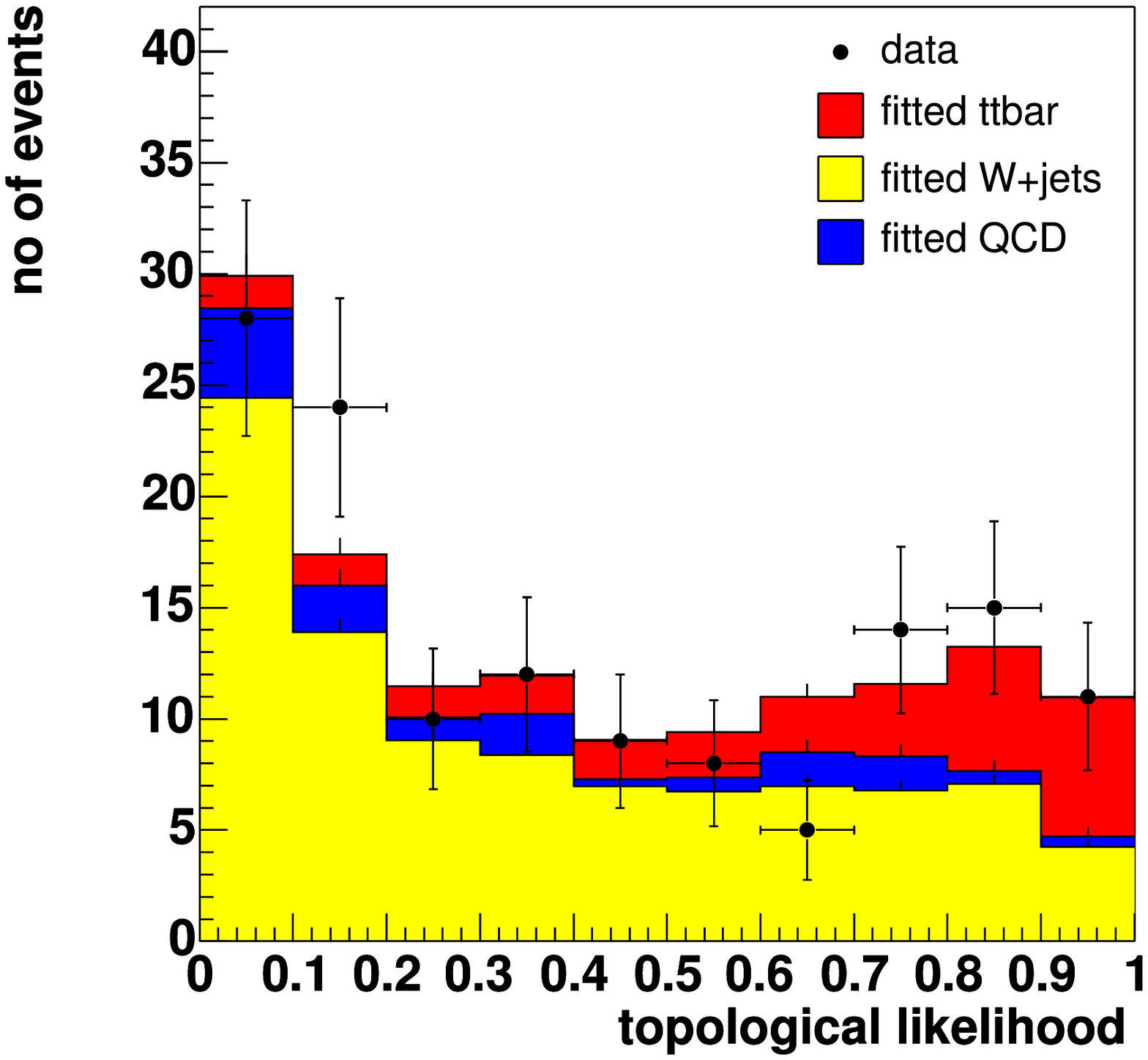}
}
\vskip -32mm
{\hskip 36mm} a) {\hskip 51mm} b) {\hskip 52mm} c)
\vskip 26mm
\caption{a) Neural Network output distribution for 
$\ell+\met+\ge3 \ jets$ 
events obtained by CDF.
b) Event likelihood discriminant distribution in 
$\mu+\met+\ge4 \ jets$ 
events obtained by D\O.
c) Event likelihood discriminant distribution in 
$e+\met+\ge4 \ jets$ 
events obtained by D\O.
\label{fig:ljets}}
\vskip -5mm
\end{figure}

CDF has used 200~pb$^{-1}$ of data to select 
$t\bar{t}\rightarrow \ell\ell+\met+\ge2\ jets$ candidates in two 
complementary analysis. In  one of the analysis
both leptons are explicitly identified as either electron or muon.
Events are selected applying basic kinematic cuts of
$p_T^{\ell+,\ell-}>20$~GeV, $\met>$25~GeV, $E_T^{j1,j2}>$20~GeV and 
total transverse energy of event $H_T>200$~GeV. 
Figure~\ref{fig:dileptons}a) shows jet multiplicity distribution
in the di-lepton events.
Number of expected $t\bar{t}$ 
signal in $\ge$2 jets bin is 10.9 with 2.7 events 
expected from background processes, while 13 candidates are observed in data. The 
measured cross-section is 
$\sigma(t\bar{t})=8.7^{+3.9}_{-2.6} (stat.) ^{+1.5}_{-1.5} (sys.)$~pb.
In another analysis, no explicit identification
is required for one of the lepton. Instead, an isolated track is considered
as a lepton candidate. Although this leads to a higher background 
contamination,
it also gives higher signal acceptance. Furthermore, the selection is 
efficient for $t\bar{t}$ events with
$W\rightarrow\tau\nu$, $\tau\rightarrow1$-$prong$ decays. 
The estimated cross-section in this analysis is  
$6.9^{+2.7}_{-2.4} (stat.) ^{+1.3}_{-1.3} (sys.)$~pb.

D\O\ has analyzed 156, 140 and 143~pb$^{-1}$ integrated 
luminosities of data to select di-electron, electron-muon 
and di-muon events.
Expected signal/background ratios in the three channels are 
$\sim$1.16, 0.32 and 3.96, respectively. 5~$e^+e^-$, 
8~$e^{\pm}\mu^{\mp}$ and 4~$\mu^+\mu^-$ events have 
been observed in data. Figure~\ref{fig:dileptons}b) shows distribution
of the  total sum of the jet transverse momenta in the selected events 
along with the
signal and background contributions. The combined 
$\sigma(t\bar{t})$ measurement for the three channels is 
$14.3^{+5.1}_{-4.3}(stat)^{+2.6}_{-1.9}(sys)$~pb.

\section{Lepton plus jets channel} 

When one of the Ws from $t\bar{t}$ 
decays leptonically and the other to hadrons, the resulting
final state contains
a high-$p_T$ lepton, large $\met$ and $\ge$3 or 4 high-$p_T$
jets. The background is dominated by
$W(\rightarrow\ell\nu)+jets$ and QCD multijet productions,
with one of the jets faking an isolated lepton. 

In 195~pb$^{-1}$ of data CDF has selected 519 events with 
$p_T^{\ell}>20$~GeV, $\met>$20~GeV and $\ge$ three jets with $E_T>$15~GeV.
In order to separate signal from the background,
various kinematic quantities have been combined
into a Neural Network discriminant, and its distribution
for the selected data events has been fitted to the 
signal and background shapes, Figure~\ref{fig:ljets}a).
The extracted 
cross-section measurement is $6.7^{+1.1}_{-1.1} (stat.) ^{+1.6}_{-1.6} (sys.)$~pb. 


\begin{figure}
\centerline{
\includegraphics[bb=10 20 570 500,width=55mm,height=60mm]
{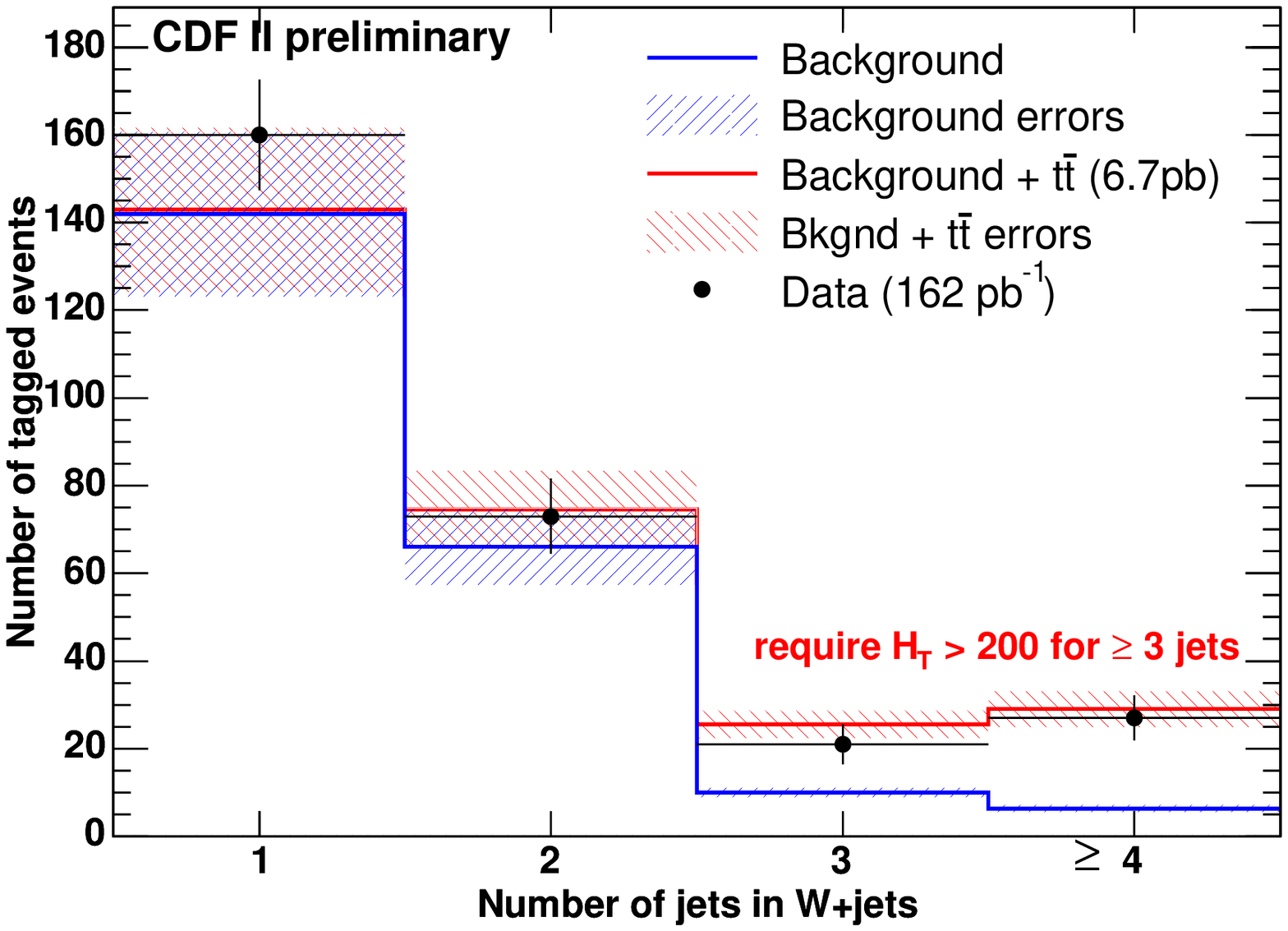}
\includegraphics[bb=10 75 570 700,width=55mm,height=60mm]
{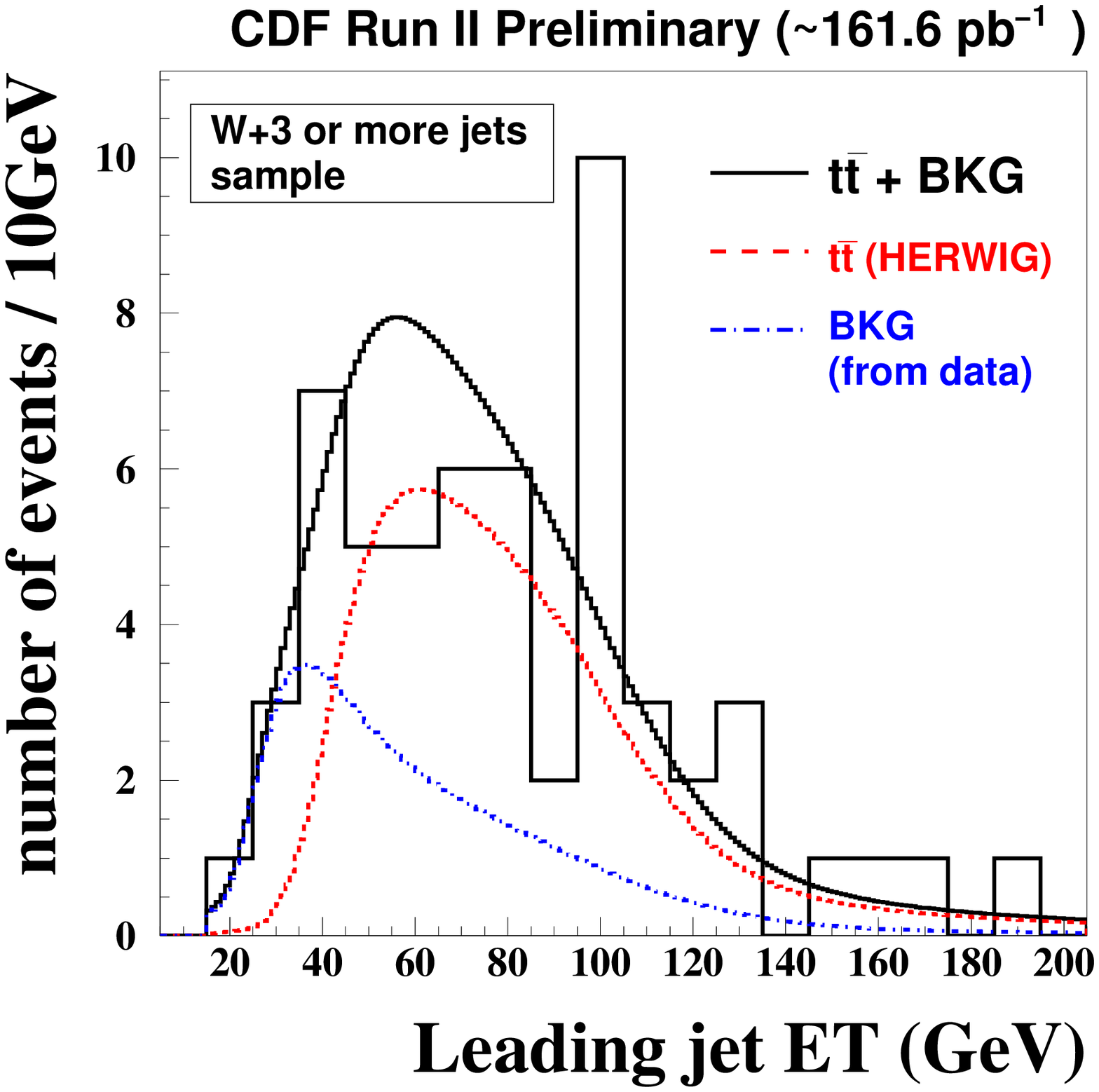}
\includegraphics[bb=10 50 570 690,width=57mm,height=60mm]
{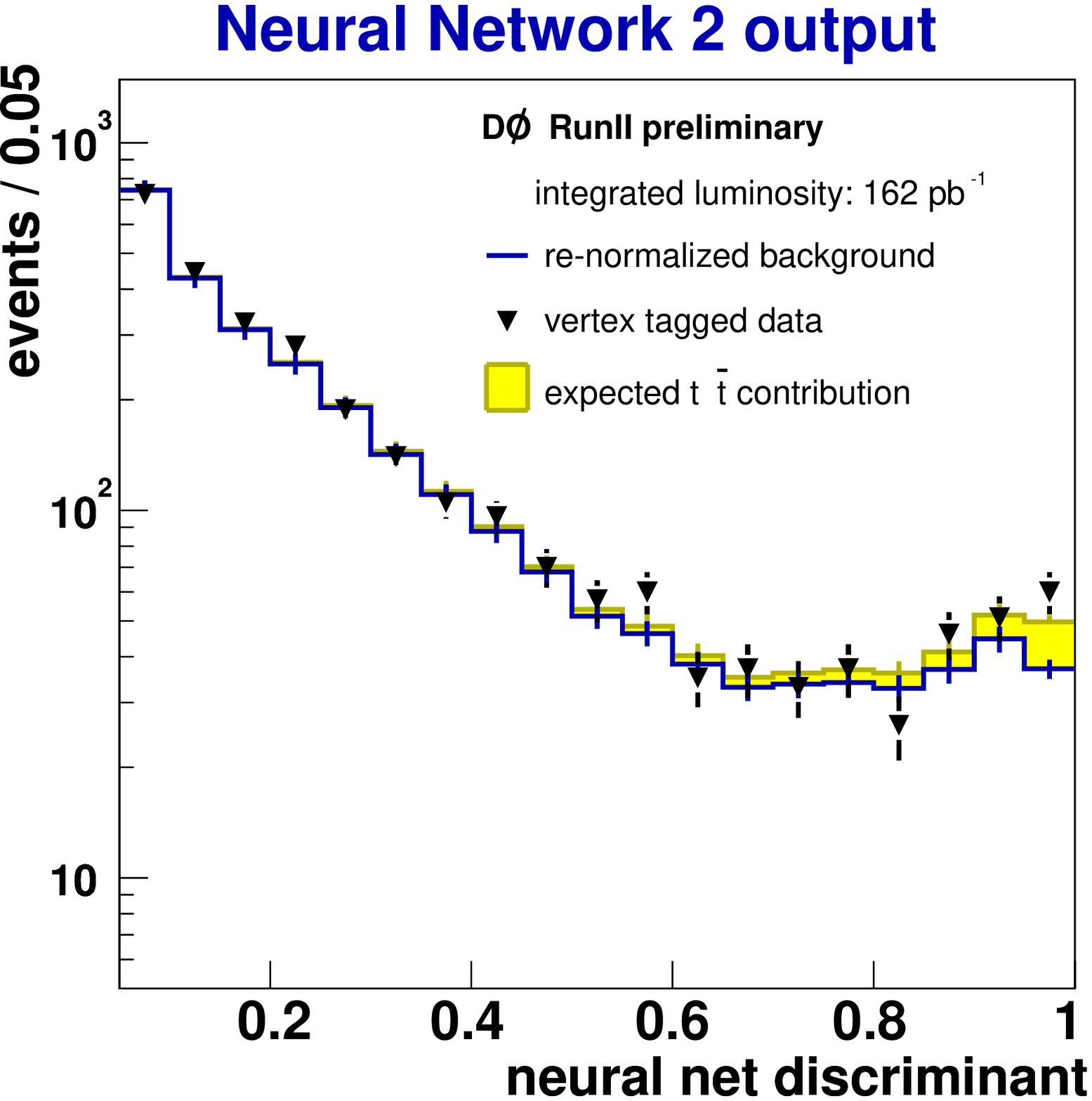}
}
\vskip -25mm
{\hskip 36mm} a) {\hskip 51mm} b) {\hskip 52mm} c)
\vskip 23mm
\caption{a) Jet multiplicity distribution obtained by CDF in $\ell+\met$ 
events with a b-tag jet. 
b) Distribution of leading jet $E_T$ in b-tagged $\ell+\met+\ge3jet$ events selected by CDF.
c) Neural Network output distribution in
$t\bar{t}\rightarrow all \ jets$ events selected by D\O.
\label{fig:btag}}
\vskip -3mm
\end{figure}

In a similar manner, D\O\ has combined topological variables
into an event likelihood discriminant.
The distribution of the discriminant for the 100 selected 
$\mu+\met+\ge4jets$ and 136 $e+\met+\ge4jets$ events are 
shown in Figure~\ref{fig:ljets}b) and c), together with the
fitted contributions from the signal and background processes. 
The analyzed event samples correspond to luminosities of
144~pb$^{-1}$, and 141~pb$^{-1}$ for the two channels, respectively. 
The cross-section is estimated to be
$7.2^{+2.6}_{-2.4} (stat.) ^{+1.6}_{-1.7} (sys.)$~pb.


\begin{figure}
\centerline{
\epsfig{figure=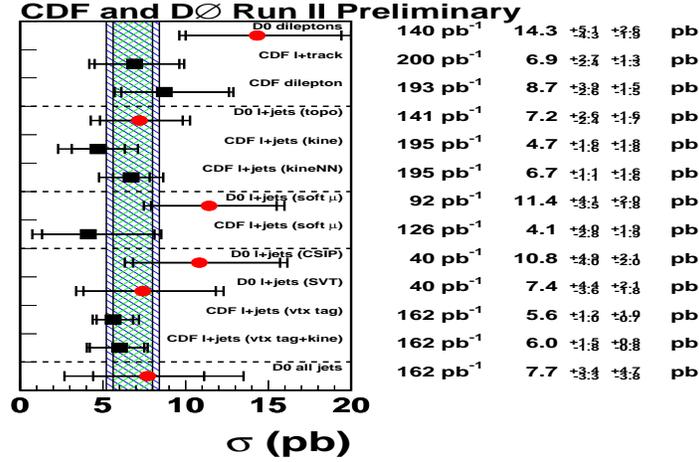,height=67mm,width=110mm}
}
\caption{Cross-section measurements for the top pair production
at the Tevatron Run~II by CDF and  D\O\ experiments in variety of
final states. The hatched bands indicated expected theoretical
cross-section. The numbers in the first column indicate used
integrated luminosities in each of the analysis.
\label{fig:summary}}
\vskip -3mm
\end{figure}


The $t\bar{t}$ decays contain two high-$p_T$ b-jets.
B-hadrons are long lived, and when coming from top 
decays, they typically travel long distance, 
$\sim$3mm,
which can be effectively resolved thanks to new 
Silicon Microvertex sub-detectors in both, CDF and  D\O\ experiments.
Applying b-tagging requirement in
$\ell+\met+\ge3jets$ events substantially reduces contamination from
$W+jets$ and $QCD$ multijet processes.
Figure~\ref{fig:btag}a) shows jet multiplicity distribution in
$\ell+\met+jets$ events obtained by CDF using 162~pb$^{-1}$ of data. 
At least 1 jet is  tagged as a b-jet. It can be seen that 
contribution from $t\bar{t}$ signal becomes dominant already in events with
3 jets. The measured cross-section in this analysis is 
$5.6^{+1.2}_{-1.0} (stat.) ^{+1.0}_{-0.7} (sys.)$~pb. 
CDF has also devised separate
analysis which makes use of event kinematic variables
in order to extract the $t\bar{t}$ cross-section in b-tagged sample
of  $\ell+\met+\ge3jets$ events.
Figure~\ref{fig:btag}b) shows distribution of the leading jet $E_T$. 
The fraction of $t\bar{t}$ events is 
extracted by fitting distribution of observed events to the 
expected shapes from signal and background processes. 
The obtained cross-section is 
$6.0^{+1.5}_{-1.8} (stat.) ^{+0.8}_{-0.8} (sys.)$~pb. 

In D\O\ two b-tag analyzes have been performed  using 40~pb$^{-1}$
of data. The analyzes use different techniques to identify displaced 
vertices from B-hadrons. They yield consistent cross-sections measurements of 
$10.8^{+4.9}_{-4.0} (stat.) ^{+2.1}_{-2.0} (sys.)$~pb. and 
$7.4^{+4.4}_{-3.6} (stat.) ^{+2.1}_{-1.8} (sys.)$~pb.
Updated analyzes with much higher luminosity 
will become available in the nearest future.

Another way to identify b-jet
is to make use of semileptonic decays of B-hadrons. 
The b-jet can be identified
by presence of `soft' muon inside a jet. 
While this is an efficient way of suppressing background, it 
also leads to relatively modest efficiency for signal events.
Both, CDF and D\O\ have performed analysis applying soft muon 
tagging, and have extracted cross-sections of 
$4.1^{+4.0}_{-2.8} (stat.) ^{+1.9}_{-1.9} (sys.)$~pb 
and 
$11.4^{+4.1}_{-3.5} (stat.) ^{+2.0}_{-1.8} (sys.)$~pb, 
respectively.

\section{All jets channel}

The most challenging of the three considered signatures
from $t\bar{t}$ events is the one  
arising from Ws decaying to hadrons, and thus leading to 
events with $\ge$6 jets. Here the dominant 
background is QCD multijet production. The only efficient way to 
suppress this background is to employ b-tagging.
D\O\ has used
secondary vertex b-tagging and combined topological variables
in 3 different Neural Networks at various stages of 
$t\bar{t}\rightarrow 6 \ jets$  analysis. 
While outputs of 
1st and 2nd  Neural Networks are used to cut on, and thereby enrich sample
in signal, 
the distribution of the 3rd NN discriminant is 
fitted to extract  fraction of the $t\bar{t}$ events.
Figure~\ref{fig:btag}c) shows the distribution of the 3rd
 Neural Network discriminant 
for the selected 220 events from  162~pb$^{-1}$ of data, together with the expected contribution
from the signal and background processes. 
The obtained cross-section  is $7.7^{+3.4}_{-3.3} (stat.) ^{+4.7}_{-3.8} (sys.)$~pb.

\section{Summary}

At the Tevatron Run~II the top pair production cross-section 
has been measured by CDF and  D\O\ Collaborations in a variety
of final states. The datasets for most
of the analysis are
larger than those at the Tevatron Run~I. The compilation of the 
results is given in Fig.~\ref{fig:summary}, along with
the recent theoretical calculations~\cite{kidonakis,cacciari} 
for $\sigma(t\bar{t})$ at $\sqrt{s}=1.96$~TeV.
The measurements, while still limited in the statistics, are consistent 
with the Standard Model expectations.




\end{document}